# Comment to the article "Possible Systematic Effects in Fomblin Coated Storage Cell Neutron Lifetime Measurements", Steve K. Lamoreaux [1]

A.P. Serebrov [a†], V.E. Varlamov [a], A.G. Kharitonov [a], A.K. Fomin [a],
Yu.N. Pokotilovski [b], P. Geltenbort [c], I.A. Krasnoschekova [a], M.S. Lasakov [a],
R.R. Tal'daev [a], A.V. Vassiljev [a], O.M. Zherebtsov [a]

[a] *Petersburg Nuclear Physics Institute, Russian Academy of Sciences, 188300 Gatchina, Leningrad District, Russia*
[b] *Joint Institute for Nuclear Research, 141980 Dubna, Moscow Region, Russia*
[c] *Institut Max von Laue-Paul Langevin, BP 156, 38042 Grenoble cedex 9, France*

## Abstract

The author of article discusses the possible systematic effects in our experiment using reference and picture from allegedly our publication. First of all this publication does not exist. The assumption of author about the probability of lower energy upscattering is in a rough contradiction (300 times) with our experimental limit which is extremely lower, about $6 \cdot 10^{-9}$ per neutron collision. The second assumption of author concerning quasi-stable orbits of UCN with higher energy due to specular neutron reflection from trap wall is also in a rough contradiction with our experimental data clear demonstrated in our previous articles as well in the last very detailed article recently presented in the arXiv.

We are very surprised that author discusses well known questions studied in our experiment in details.

[†] A.P. Serebrov, Petersburg Nuclear Physics Institute,
Russian Academy of Sciences,
188300 Gatchina, Leningrad Region, Russian Federation
Tel. (7-81371) 46 001
E-mail: serebrov@pnpi.spb.ru

In the article [1] the possible sources of systematic errors are considered which could take place in two neutron lifetime experiments in when the fully fluorinated polymers were used for coating the surface of the ultracold neutron storage cells. In the first part of the article the author considers the experiment of W. Mampe et al. [2] carried out with the storage chamber covered with liquid Fomblin at room temperature. The author did not find any possible systematic error due to uncontrolled quasi-elastic scattering on the surface of liquid Fomblin in the experiment [2] but had come to assumptions about possible systematic error in our experiment in which similar polymer was used as storage chamber wall cover in solid state at low temperature [3].

In our experiment the effect of quasi-elastic upscattering has been carefully studied as a function of temperature. It was shown that this effect is radically suppressed at the temperature $-160°C$, at which our neutron lifetime measurements was performed. Nevertheless the author of the article [1] insists that in our experiment the effect of quasi-elastic scattering has been observed and demonstrated the Fig.5 from allegedly our publication: P. Geltenbort, A. Serebrov, V. Varlamov, A. Kharitonov, R. Tal'dayev, O. Zherebtsov, B. Yerozolimsky, N. Achiwa, A. Pichlmaier, K. Schreckenbach, O. Kwon, and A. Steyerl, 11-th Internat. Seminar on Interaction of Neutrons with Nuclei (ISINN 11), Dubna, May 25-28, 2003.

However, this publication does not exist. There is no such contribution in the Conference proceedings and it was never presented at this Conference. Most probably this Fig. 5 in [1] is taken from our logbook, it contains the count rate data before subtraction of background effect due to neutron background in reactor hall and due to some part of UCN still leaking because the time of monitoring is not enough. The author of the article [1] interpreted these intermediate results as an indication of significant quasi-elastic upscattering neutron loss during neutron storage, inspite of statement of our works.

We are very grateful to the author of this article [1] for his interest to our intermediate results and his analysis of our experiment. But we are absolutely sure that only published results can be the subject of discussion and especially of negative-like comments in scientific literature.

In connection with this situation we presented in the arXiv detailed description of our experiment [4]. This article is the extended version of our first preprint [5], which was prepared before our publication [3] in Physics Letters B. This preprint has been distributed rather widely. The same detailed information has been published in [6] and has been repeatedly reported at the Conferences, for example [7,8].

In our publication [4] it was shown that the probability of low energy upscattering of UCN is less than $6·10^{-9}$ per neutron collision with the wall if the energy transfer is about 20 neV. The upper limit of correction for the neutron lifetime value in this case is 0.03 s, i. e. 30 times less than the experimental error and 300 times less than the conclusion of [1].

In addition, in our publications [3-6] it has been shown that there is no effect of quasi-stable orbits for UCN with higher energies due to specular neutron reflections from the walls. Nevertheless the author of [1] reminds again about the possibility of this type of effect in our experiment and even recommends to repeat experiment with another trap surface. We appreciate very much this recommendation and repeat that the level of specular reflections was already estimated from our present experiment. We recommend to author of article [1] to read more attentively our articles. Sure, the new measurements of the neutron lifetime are very important by different experimental groups with different experimental methods.

Lastly, it should be mentioned, that the statement in [1] that quasi-elastic scattering on the liquid Fomblin does not bring a systematic correction in experiment [2] is in contradiction to our detailed Monte Carlo simulation of experiment [2]. These calculations were reported [9] at 5-th International UCN Workshop "Ultracold and Cold Neutrons. Physics and Sources", Peterhof, Russia, 13-18 July 2005. In the work [9] it was shown that correction for the neutron lifetime from the experiment [2] is 2.5 s instead of 9 s, which was used in article [2]. Therefore new corrected value from experiment [2] will be 881.1±3.0 s, that is in agreement with our result 878.5±0.8 s. Additionally, it should be mentioned that the effect of quasi-elastic scattering is most dangerous when the initial neutron spectrum has the upper cutoff higher than the boundary energy of the Fomblin wall. In this case the regular leakage of super-barrier neutrons through the wall takes place from the first moment of storage process. In the case when there is the gap between upper cutoff of the initial neutron spectrum and the boundary energy of wall as it was in experiment MAMBO II [10], the effect of UCN leakage through energy barrier would appear only at longer neutron holding times. Therefore the correction is considerably lower. It is supported by the experiment MAMBO II [10] which used neutron spectrum with the energy cutoff below the Fomblin boundary energy, and in result the neutron lifetime value 881±3 s has been obtained. This value is in reasonable agreement with our result [3] 878.5±0.8 s as well as with the corrected result of the experiment [2]: 881.1±3.0 s. Thus the main cause of the discrepancy is in significant quasi-elastic upscattering from the liquid Fomblin in result of neutron interaction with viscoelastic surface modes, and in the initial neutron spectrum spreading higher than the boundary energy of the wall's storage chamber.

In conclusion we would like to thank Steve K. Lamoreaux who had urged us to publish more detailed article of our experiment.